# High-throughput Search for Metallic Altermagnets by Embedded Dynamical Mean Field Theory

*Xuhao Wan [a], Subhasish Mandal [b], Yuzheng Guo [c], Kristjan Haule [a] ***

[a] Center for Materials Theory, Department of Physics and Astronomy, Rutgers University, Piscataway, NJ 08854, USA

[b] Department of Physics and Astronomy, West Virginia University, Morgantown, WV 26506, USA

[c] The Institute of Technological Sciences, Wuhan University, Wuhan, 430072, China

* Corresponding Author: k.haule@physics.rutgers.edu (Kristjan Haule)

**ABSTRACT**

Altermagnets (AM) are a novel class of magnetic materials with zero net magnetization but broken time-reversal symmetry and spin-split bands exceeding the spin-orbit coupling scale, offering unique control of individual spin-channel and high charge-spin conversion efficiency for spintronic applications. Still, only a few metallic altermagnets have been identified, and discovering them through trial-and-error is resource-intensive. Here, we introduce a high-throughput screening strategy to accelerate the discovery of materials with altermagnetic properties. By combining density functional theory (DFT) with embedded dynamical mean-field theory (eDMFT), our approach improves the accuracy in predicting metallicity and spin splitting, especially in transition-metal-rich compounds. An automated workflow incorporates pre-screening and symmetry analysis to reduce both human effort and computational cost. This approach identified two previously unreported metallic altermagnets, CrSe and $CaFe_4Al_8$ (in addition to two known altermagnets, CrSb and $RuO_2$), as well as a dozen semiconducting altermagnets among over 2,000 magnetic materials. Our findings reveal that while altermagnets are abundant among magnetic materials, only a tiny fraction is metallic.

# INTRODUCTION

An altermagnet (AM) is a novel type of magnetic material, distinct from traditional ferromagnet or antiferromagnet (AFM).[1] Its defining feature is a magnetic order with vanishing total magnetization and non-relativistic spin-split bands.[2] While the magnetic ordering in real space resembles that of collinear antiferromagnetism, the spin-splitting of bands in momentum space represents a natural characteristic of ferromagnetism.[2,3] Unlike antiferromagnets, altermagnets exhibit spin polarization that alternates between different crystal sublattices in more complex symmetries, cancelling the overall magnetic moment.[1] This results in a unique electronic structure with spin-selective band splitting that does not rely on relativistic effects such as spin-orbit coupling. Moreover, the magnitude of the splitting in altermagnets can significantly exceed the spin-orbit interaction scale. In contrast to antiferromagnets, the spin-splitting of bands in AM allows for the control of individual spin channels.[4] Simultaneously, altermagnets exhibit superior charge-spin conversion efficiency, such as current-induced magnetization, the switching can be faster and more energy efficient, and AM does not suffer from stray fields inherent to ferromagnets. In this context, they are particularly promising for spintronic applications[4,5] and as active elements in memory and computing devices.[6] Despite these promising properties, the number of known metallic altermagnets remains limited, necessitating the development of robust methodology for their discovery.

In recent years, the study of altermagnets has garnered significant attention as it is debated to be a new phase of matter. Several materials have already been reclassified as altermagnets, including thin films of rutile metal $RuO_2$,[7] which show anomalous Hall effect,[8,9] charge-spin conversion,[10] and spin-torque phenomena,[11] MnTe in bulk, which shows lifted Kramers spin degeneracy,[12] and CrSb thin films with large band splitting.[13] The metallic nature of $RuO_2$ and CrSb allows for efficient charge transport, further enhancing their relevance in spintronics and electronic devices. However, discovering new altermagnets with large splitting through serendipitous search is inherently challenging and resource-intensive. To address this, high-throughput screening with reliable first-principles calculations has emerged as a promising alternative. As shown in **Fig. 1a**, the strategy combines properties from the materials database and first-principles calculations to screen all candidate materials simultaneously, ultimately identifying promising altermagnets that meet the necessary conditions for applications. The stepwise screening process ensures the identification of promising metallic altermagnets at a lower cost and in a shorter time. However, the limitations of standard density functional theory (DFT) in capturing strong electron correlations, arising from transition metal ions hosting electrons that are neither fully localized nor fully itinerant, hinder the efficient classification of materials into metals and insulators, and limit predictions of spin splitting without fine-tuning of the Coulomb U parameter.[14,15] Thus, using computational methods that go beyond standard DFT is important.

Metallic magnetic materials are particularly challenging to describe due to the coexistence of orbitals forming wide and narrow bands, often resulting in collective electronic behaviour.[16] The beyond-DFT method based on hybrid functionals is known to be inefficient for describing metals,[17] making the computational search for metallic altermagnets more challenging. In this context, Dynamical Mean-Field Theory (DMFT) offers a significant advantage, as it can describe strong local correlations on a given site exactly, with longer-range correlations handled in a mean field way.[18,19] Unlike conventional DFT or DFT+U approaches, DMFT explicitly accounts for local quantum and valence fluctuations on a correlated transition metal ion, leading to dynamical correlations and short quasiparticle lifetimes away from the Fermi level.[18-20] In this way, DMFT is expected to provide a more accurate electronic structure for altermagnets at the boundary between localized and itinerant behaviour.[21-23] In the present work, the high-throughput calculations (HTC) screening framework based on DFT+embedded DMFT (DFT+eDMFT)[24,25] is developed and applied to identify new altermagnets with large spin-split bands efficiently. The subsequent analysis provides insights into the discovery of new altermagnets via high-throughput screening within more extensive material databases, paving the way for the design of next-generation spintronic devices.

**RESULTS**

Conventional Néel antiferromagnets have a vanishing total magnetic moment and exhibit Kramers' degeneracy of quasiparticle bands throughout the reciprocal momentum space in the absence of spin-orbit coupling (**Fig. 1b**).[1,4] Consequently, the electrical behaviour of antiferromagnets often resembles that of nonmagnetic materials. In contrast, altermagnets are characterized by the absence of Kramers' degeneracy in some parts of momentum space, where the quasiparticle bands are spin polarized even in the absence of spin-orbit coupling. Nevertheless, the total magnetic moment still vanishes. The vanishing of the moment is protected by specific crystallographic symmetries of the magnetic ground state in altermagnets that have no ferrimagnetic component.[1] Specifically, if the vanishing of the total magnetic moment (M=0) is protected by the combined symmetry of inversion (P) and time-reversal (T), the Kramers' degeneracy at each momentum point is inevitable, and the material should be classified as conventional AFM. Similarly, if M=0 is protected by the combination of time-reversal (T) and translation (t), the degeneracy is unavoidable in the absence of spin-orbit coupling or in a centrosymmetric space group, resulting in conventional antiferromagnet (**Fig. 1b**).[12] In contrast, as shown in **Fig. 1c**, the vanishing moment of altermagnet is protected by proper or improper rotation or non-symorphic operation (but not inversion or translation).

In colinear magnetic systems, the symmetry operations of the space group **G** can be decomposed into two categories: i) operations that preserve the orientations of the magnetic moments and interchange atoms within one of the two spin sublattices, denoted by **H**, which are a subgroup of **G**, and ii) the coset A**H** = **G**−**H** generated by transformations A that flip the magnetic moments, which can also be accompanied by any operation from **H** that preserves the orientations of the moments. The total space group is hence decomposed into **G**=**H**+A**H**. If **G** = **H**, which means all operations in the system preserve the spin configuration, the material is ferromagnet or ferrimagnet. If A***H** contains the combined inversion and time-reversal symmetry (P*T) or combined translation and time-reversal symmetry (t*T), the material is conventional AFM, as previously discussed. Otherwise, it is an altermagnet candidate.[1,26] This algorithm is also consistent with the previous classification,[26,27] but here, we also consider non-collinear magnets, which can become collinear when spin-orbit interaction is switched off.[28] Similarly, some ferrimagnetic materials with vanishing magnetic moments can become altermagnets in the absence of SO;[28] hence we extended the pool to include these candidates. An example is CrSe, which corresponds to a ferrimagnetic space group, but as we will show below, its collinear analog has a typical property of an altermagnet. Thus, we classify CrSe as altermagnet, which becomes non-collinear in the presence of SO coupling.

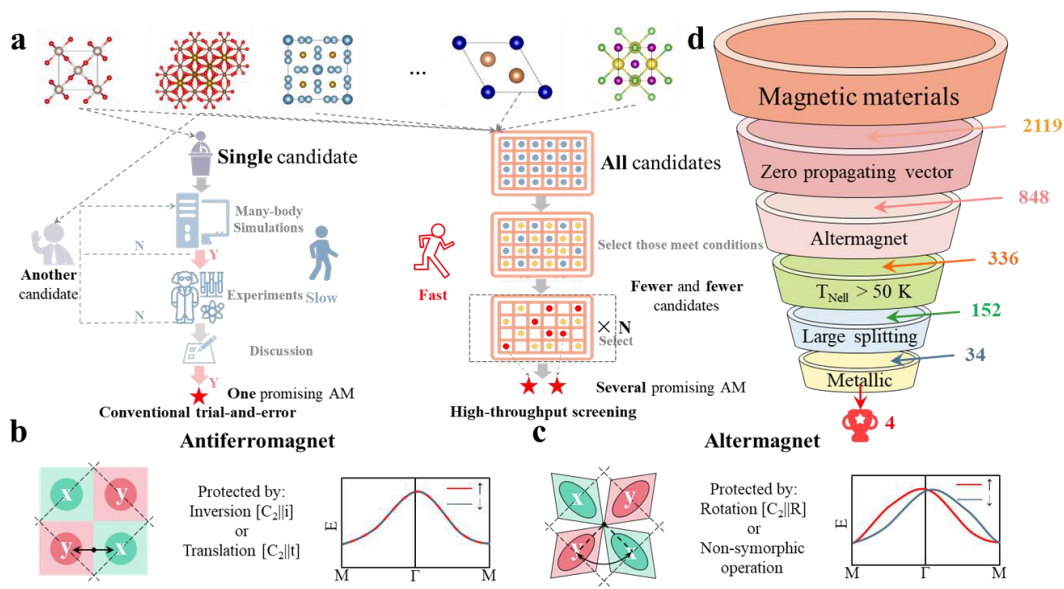

**Figure 1 The illustration of the high-throughput screening strategy and the comparison between antiferromagnet and altermagnet. a,** the schematic comparison of the conventional experimental trial-and-error method and the high-throughput screening method in discovering AM. **b,c,** the difference between an antiferromagnet and an altermagnet. **d,** The flow chart for screening large band-splitting metallic altermagnet in this study.

Using the known characteristics of altermagnet, it is possible to design the high-throughput screening workflow, shown in **Fig. 1d**. The initial set of candidate materials is obtained from the Bilbao MAGNDATA database,[29] which is an open database dedicated to magnetic materials. It contains over 2,000 crystalline materials for which the magnetic configuration and crystal structure are experimentally determined. Its comprehensiveness, reliability, and standardized format make it an excellent choice for high-throughput screening. For the initial screening, we selected 1695 magnetic materials, in which the crystal structure is not disordered. Of those, 848 materials have a 0-propagation vector, 758 have a single propagating vector, and 89 have two propagating vectors. In the second step, we analysed magnetic crystal structure using the *pymatgen* package[30] and determined if symmetries of the space group allow the altermagnetic splitting and vanishing total magnetic moment. For example, we checked that t*T and P*T symmetry are absent, including the possibility that the inversion center is at the arbitrary point in space. The specific steps are shown in **Fig. 2a**. Ultimately, 289 materials with 0-propagating vector, 33 with single, and 14 with two propagating vectors remained as potential altermagnetic candidates.

Next, we screen these 336 potential altermagnets based on their Néel temperatures, with a threshold of higher than 50 K. This is because most potential applications of altermagnet require materials to operate at temperatures close to the room temperature.[31,32] The Néel temperatures of all materials are directly obtained from the MAGNDATA database. This screening resulted in 152 potential altermagnets with sufficiently high Néel temperatures.

Next, we perform t-distributed Stochastic Neighbour Embedding (t-SNE) analysis of all 2119 magnetic materials to visualise the clustering of altermagnetic candidates within the space of all magnets. t-SNE is a nonlinear dimensionality reduction technique commonly used for the visualization of high-dimensional data.[33] Its primary function is to map high-dimensional data into low-dimensional space to enable intuitive visualization of patterns and structures within the data. It is particularly suitable for handling nonlinear, complex datasets, making it widely applicable in high-throughput materials screening. Herein, we comprehensively consider the fundamental physicochemical properties of magnetic materials, including elemental composition, crystal system, lattice constants, and propagation vectors. One-hot encoding is applied to the elemental composition and crystal systems to ensure the independence of elements and crystal systems, avoiding order bias. The input details for t-SNE analysis can be found in **Supplementary Note 1**. After reducing the high-dimensional features of all materials to two dimensions, the visualized results are shown in **Fig. 2b**.

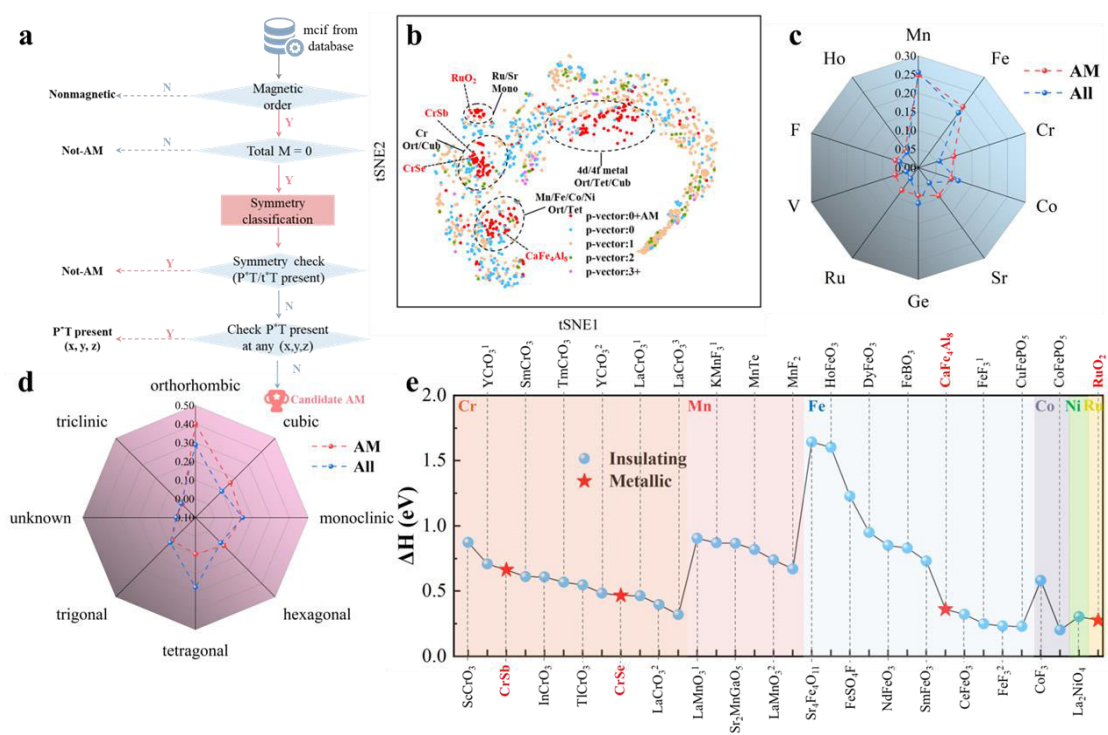

**Figure 2 The results and analysis of the high-throughput screening process. a,** the schematic of the method to check if a material is altermagnet. **b,** the results of the t-SNE analysis. **c-d,** the radar chart of the proportion of the magnetic element and crystal system in all magnetic materials and altermagnet candidates. **e,** the results of the band splitting and metallic properties of the four metallic altermagnets and top 30 insulating altermagnets with significant band splitting.

The t-SNE analysis results show that when dividing all magnetic materials into five categories based on their propagation vectors and whether they are altermagnetic, including 0 propagation vector altermagnet, 0 propagation vector non-altermagnet, 1 propagation vectors, 2 propagation vectors, and 3 or more propagation vectors, only the altermagnet category exhibits significant clustering. This indicates that altermagnets share a distinctive type of underlying characteristic that makes them more homogeneous than other magnetic materials. This might be due to their unique magnetic symmetry, which governs the spatial arrangement of their magnetic moments. Conversely, other ferromagnets, antiferromagnets, ferrimagnets, *etc.*, do not exhibit any clustering and are distributed mostly randomly due to the more diverse nature of their magnetic behaviours. Furthermore, the visual inspection shows that altermagnetic clusters can be subdivided into four smaller clusters related to their elemental composition and crystal symmetry. To confirm that, we conduct a statistical analysis and annotate the plot in **Fig. 2b** with the elemental compositions and crystal symmetries within each cluster. The resulting four altermagnetic groups are: i) the lower left group is dominated by the 3d transition metal ions: Mn, Fe, Co, and Ni, and the orthorhombic and tetragonal crystal systems; ii) the second is dominated by Cr ions and the orthorhombic and cubic crystal systems; iii) the third is dominated by Ru and Sr ions and the monoclinic crystal system; iv) and the fourth group is dominated by 4d or 4f ions and the orthorhombic, tetragonal, or cubic crystal systems. This suggests that only a few metal ions (most are magnetic ions simultaneously) in their respective crystal symmetries dominate the space of altermagnets.

We analyze the abundance of each metal ion in all magnets and altermagnets. **Fig. 2c** shows that Mn is the most common magnetic ion. However, it is not more likely to be altermagnet than conventional AFM or ferromagnet. On the other hand, Cr, Ru, and Fe ions are more often AM than an average magnetic material. Sr is not a magnetic ion, but it plays a role in structural stabilization and electronic modulation in AM therefore it is commonly found in AM.

We also determine which crystal structures are most common among magnets and altermagnets (**Fig. 2d**). Notably, magnetic materials with orthorhombic and cubic crystal symmetry are more likely to be altermagnets. In contrast, those with tetragonal crystal symmetry are less likely to be altermagnets. This is because crystals with higher symmetry can have more symmetry operations that protect the vanishing of the total magnetic moment while simultaneously breaking Kramers degeneracy. On the other hand, altermagnets are very rarely found in tetragonal symmetry due to the presence of a fourfold rotational axis, which more strongly constrains the arrangement of magnetic moments, making it more difficult to form the anisotropic spin configuration required for altermagnets. This insight suggests that future design of altermagnets, using more extensive databases of known materials should focus on materials containing Cr, Fe, and Mn, and with an emphasis on the orthorhombic and cubic crystal symmetry, but not triclinic or tetragonal symmetry.

In the final stage of the screening process, we determine which material within the 152 potential altermagnetic candidates with high enough Neel temperature is metallic and has considerable spin-splitting of the bands around the Fermi level. To that end, we use the DFT+DMFT many-body method to provide more accurate electronic structure simulations. All calculations are performed with the eDMFT package[24,25] which implements the exact double-counting between DFT and DMFT[34], and optimizes well-defined many-body functional that account for correlations local to each magnetic ion exactly. We did not include spin-orbit coupling which the definition of altermagnets does not require. We obtain the spectral functions of all 152 potential altermagnets, which reveals that most potential altermagnets are insulators, except for just a handful of metallic materials. We then compared the magnitude of the band splitting of these altermagnets. The measure of the spin splitting (ΔH) is calculated by integrating the first moment in frequency of the difference between spin-up and spin-down spectral functions from -3 eV to 3eV around the Fermi level, filtering out contributions from k-points with negligible splitting, and averaging the results over remaining momentum points. The detailed formula of ΔH is available in **Supplementary Note 2**. The results are shown in **Fig. 2e**, where we plot the four metallic altermagnets with significant band splitting, including CrSb, CrSe, $CaFe_4Al_8$, and $RuO_2$, and the top thirty insulating altermagnets with considerable band splitting (ΔH > 0.2 eV). They are categorized by the magnetic atom, including Cr, Mn, Fe, Ni, Ru, and Os.

After the high-throughput screening process, most potential altermagnets with band splitting ($\Delta_H$ > 0.01 eV) are listed in **Table 1**, along with their MAGNDATA database ID, band gap (if insulating), local Coulomb repulsion U estimated for each material, theoretical spin moment (without orbital moment t-M), experimental magnetic moment (e-M), ΔH, and Néel temperature ($T_{Néel}$). Four metallic altermagnets with significant band splitting include CrSb, CrSe, $CaFe_4Al_8$, and $RuO_2$. Among them, CrSb and $RuO_2$ have recently been reported in experimental studies,[9,13] supporting the validity of our high-throughput screening strategy. **Table 1** indicates that metallic altermagnets are extremely rare. Among the dozens of altermagnets exhibiting apparent band splitting, only five materials possess metallic properties. The remaining altermagnets are insulating. Among them, $SrFe_4O_{11}$ stands out due to its simplicity of the Fermi surface when hole doped. Notably, CrSe possesses two distinct propagation vectors.

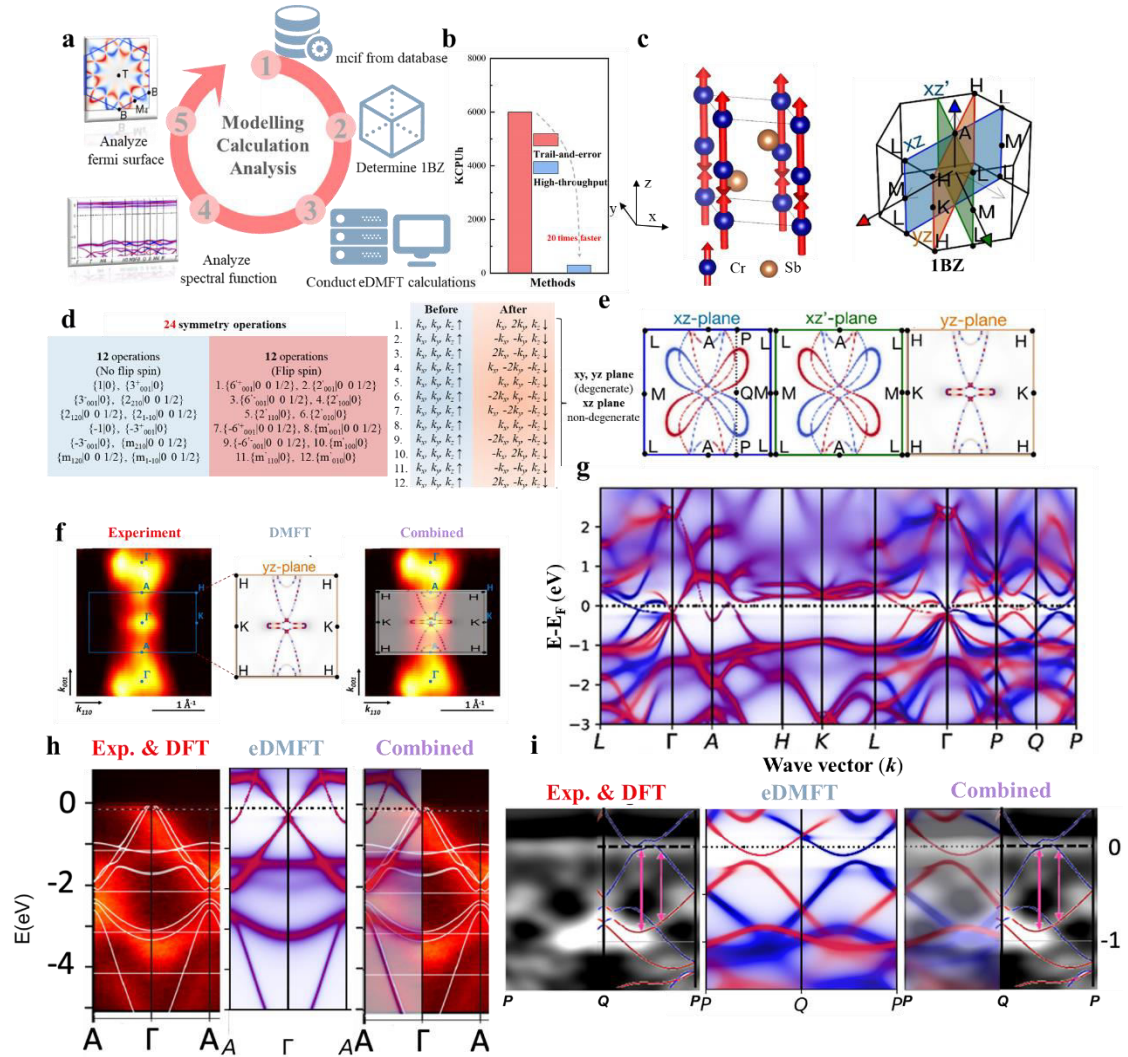

**Figure 3 The eDMFT calculations for the discovery of potential altermagnet. a,** the schematic of screening altermagnets by eDMFT calculations. **b,** the comparison of research

efficiency between traditional trial-and-error and high-throughput methods. **c,** the crystal structure and the first Brillouin zone of CrSb. **d,** the symmetry operation analysis of CrSb for determining which fermi surface is degenerate as potential altermagnet. **e,** the xz, xz', and yz fermi surface of CrSb from eDMFT calculations where minority and majority electrons are marked red and blue, respectively. **f,** the comparison of the yz fermi surface of CrSb with its experimental SX-ARPES in the *Γ-K-A* plane (yz plane).[13] **g,** the spectral function of CrSb along the high-symmetry path by eDMFT calculations. **h,** the comparison of the spectral function of CrSb with the experimental SX-ARPES intensity along the high symmetry *Γ-A* path with p-polarized photons.[13] **i,** the comparison of the spectral function of CrSb with the experimental SX-ARPES intensity along the high symmetry *P-Q* path with p-polarized photons.[13]

The specific workflow of performing eDMFT calculations to determine potential altermagnetism is presented in **Fig. 3a**. The process begins with obtaining the mcif file from the database, followed by determining the correlated magnetic ions and their symmetry relations, finding the best-suited local axis on each correlated ion that preserves the symmetry of the space group, and producing all input files for eDMFT calculation. The eDMFT simulation produces the exchange splitting through the self-energy on all magnetic ions, and the spectral function and the charge density of the solid. Further analysis reveals the size of the AM splitting and its usefulness for application. The efficiency of the high-throughput screening strategy with the conventional trial-and-error method is shown in **Fig. 3b**. An eDMFT calculation for a single material takes between 600 to 6000 CPU hours (CPUh). For studying all 2,000 magnetic materials in the database, 6 million CPUh would be required. In contrast, the high-throughput screening workflow requires only 300 thousand CPUh, yielding a 20-fold efficiency improvement.

Then, taking CrSb as an example, the specific workflow to perform calculations and analyze results is detailed. The crystal structure and the first Brillouin zone of CrSb are shown in (**Fig. 3c**), with magnetic moments localized on Cr atoms with magnitude 2.84 μB and P6$_3$'/m'm'c magnetic space group. There are 24 symmetry operations associated with this space group: twelve operations that do not flip spin (e.g.,{-1|0}, {3$^+_{001}$|0}), belonging to **H** subgroup, and twelve operations that flip spin (e.g., {2'$_{001}$|0 0 1/2},{m$_{1-10}$|0 0 1/2}) and are part of **A$^*$H**. Analysis of the **A$^*$H** operations reveals that the time reversal symmetry in the xy and yz plane is protected by {2'$_{001}$|0 0 1/2} E(***k$_x$, k$_y$, 0***; ↑) → E(***k$_x$, k$_y$, 0***; ↓) and {m$_{1-10}$|0 0 1/2} E(***0, k$_y$, k$_z$***; ↑) → E(***0, k$_y$, k$_z$***; ↓). At the same time, the xz plane is not degenerate as E(***k$_x$, 0, k$_z$***; ↑) will not translate to E(***k$_x$, 0, k$_z$***; ↓) by any symmetry operation. Therefore, as shown in **Fig. 3e**, the eDMFT fermi surface confirms the degeneracy and splitting of the yz and the xz planes, respectively. Due to symmetry constraints, all three equivalent xz planes show strong time-reversal splitting, with 60° (120°) rotation having the opposite (same) splitting of bands. Moreover, the comparison of our predicted yz fermi surface with the experimental Soft X-ray Angle-resolved Photoemission Spectroscopy (SX-ARPES) results shows a good agreement (**Fig. 3f**).[13] The analysis also reveals large splitting in one of the initially reported altermagnets RuO$_2$, with the results presented in **Supplementary Figure 1**. We note that the eDMFT calculations of RuO$_2$ are also consistent with the ARPES experiments.

The spectral function of CrSb along the high-symmetry path indicates there is a strong splitting between the *L-Γ* line, consistent with the above analysis (**Fig. 3g**). Furthermore, we compared the spectral function of CrSb with the experimental SX-ARPES intensity.[13] In **Fig. 3h**, the eDMFT spectral function along the *Γ-A* path matches well with the experimental high intensity area along the *A-Γ-A* path, and is not very different from DFT (white lines). Similarly, the eDMFT spectra along the $P$-$Q$ path is in slightly better agreement with the experiment than DFT (red/blue lines), validating the reliability and effectiveness of the eDMFT calculations (**Fig. 3i**).

**Table 1 The MAGNDATA list of interesting altermagnets:** MAGNDATA ID, BNS magnetic space group,[29] band gap, Coulomb U value, theoretical spin moment (t-M),[29] experimental magnetic moment (e-M),[29] band splitting ($\Delta H$), and $T_{Nell}$[29] of altermagnets with substantial band splitting obtained from the high-throughput screening. '/' means unknown moment. FM implies that the material is not strictly altermagnet but is weak ferrimagnetic with finite splitting at $\Gamma$ point. Two metallic altermagnets with minimal band splitting are also included.

| Material | ID | Mag. Space group | Gap/eV | U/eV | t-M/$\mu_B$ | e-M/$\mu_B$ | $\Delta H$/eV | $T_{Nell}$/K |
|---|---|---|---|---|---|---|---|---|
| CrSb | 0.528 | $P6_3'/m'm'c$(#194.268) | metal | 6 | 2.84 | 2.50 | 0.66 | >600C |
| CrSe | 2.35 | $P31m'$(#157.55) | metal | 6 | 3.50 | 3.49 | 0.47 | 300C |
| CaFe$_4$Al$_8$ | 0.236 | $I4'/mmm'$(#139.535) | metal | 6 | 0.88 | 0.71 | 0.36 | 180 |
| RuO$_2$ | 0.607 | $P4_2'/mmm'$(#136.499) | metal | 6 | 1.20 | / | 0.28 | >300 |
| SrMnSb$_2$ | 0.768 | $Pn'a'2_1$(#33.148) | metal | 6 | 3.51 | 3.78 | 0.01 | 304 |
| Cr$_2$S$_3$(FM) | 0.5 | $P-1$(#2.4) | 0.75 | 8 | 2.90 | 1.20 | 3.10 | 122 |
| CaFe$_2$O$_4$(FM) | 0.968 | $Pn'ma'$(#62.448) | 2.00 | 8 | 3.95 | 4.26 | 2.52 | 293 |
| Sr$_4$Fe$_4$O$_{11}$ | 0.402 | $Cmm'm'$(#65.486) | 0.40 | 6 | [3.60, 4.00] | [0, 3.52] | 1.64 | 232 |
| HoFeO$_3$ | 0.991 | $Pn'ma'$(#62.448) | 2.30 | 8 | 4.40 | 4.60 | 1.60 | 647 |
| FeSO$_4$F | 0.128 | $C2'/c'$(#15.89) | 3.00 | 8 | 4.60 | 4.30 | 1.23 | 100 |
| BiCrO$_3$[1](FM) | 0.139 | $P-1$(#2.4) | 1.10 | 8 | 2.50 | 1.70 | 1.14 | 80 |
| BiCrO$_3$[2](FM) | 0.138 | $C2/c$(#15.85) | 1.10 | 8 | 2.50 | 2.04 | 0.98 | 114 |
| DyFeO$_3$ | 0.836 | $Pn'ma'$(#62.448) | 2.20 | 8 | 4.40 | 3.60 | 0.95 | 650 |
| LaMnO$_3$[1] | 0.1 | $Pn'ma'$(#62.448) | 0.80 | 8 | 3.88 | 3.87 | 0.90 | 139.5 |
| ScCrO$_3$ | 0.307 | $Pnma$(#62.441) | 2.43 | 8 | 2.80 | 2.70 | 0.87 | 73 |
| KMnF$_3$[1] | 0.433 | $I4/mcm$(#140.541) | 2.60 | 8 | 4.56 | / | 0.87 | 86.8 |
| Sr$_2$MnGaO$_5$ | 0.823 | $Im'a2'$(#46.243) | 1.40 | 8 | 3.75 | 3.16 | 0.87 | 180 |
| NdFeO$_3$ | 0.336 | $Pn'ma'$(#62.448) | 2.40 | 8 | 4.40 | 3.84 | 0.85 | 760 |
| FeBO$_3$ | 0.112 | $C2'/c'$(#15.89) | 3.50 | 8 | 4.60 | 4.70 | 0.83 | 348 |
| MnTe | 0.800 | $Cmcm$(#63.457) | 0.20 | 6 | 4.60 | 4.60 | 0.82 | 323 |
| LaMnO$_3$[2] | 0.642 | $Pn'ma'$(#62.448) | 0.70 | 8 | 3.40 | 3.70 | 0.74 | 323 |
| SmFeO$_3$ | 0.380 | $Pn'ma'$(#62.448) | 1.90 | 8 | 4.40 | 2.80 | 0.73 | 670 |
| YCrO$_3$[1] | 0.586 | $Pn'ma'$(#62.448) | 3.00 | 8 | 2.76 | 2.96 | 0.71 | 141 |
| MnF$_2$ | 0.15 | $P4_2'/mnm'$(#136.499) | 2.90 | 10 | 4.60 | 4.60 | 0.67 | 67 |
| SmCrO$_3$ | 0.696 | $Pn'ma'$(#62.448) | 1.40 | 8 | 2.27 | 2.72 | 0.61 | 192 |
| InCrO$_3$ | 0.308 | $Pnma$(#62.441) | 0.80 | 6 | 2.90 | 2.50 | 0.61 | 93 |
| CoF$_3$ | 0.334 | $R-3c$(#167.103) | 2.60 | 8 | 3.66 | 3.20 | 0.58 | 460 |
| TmCrO$_3$ | 0.587 | $Pn'ma'$(#62.448) | 3.00 | 8 | 2.74 | 2.58 | 0.57 | 124 |
| TlCrO$_3$ | 0.309 | $Pnma$(#62.441) | 0.10 | 8 | 2.80 | 2.46 | 0.55 | 89 |
| YCrO$_3$[2] | 0.947 | $Pn'ma'$(#62.448) | 3.00 | 8 | 2.74 | 2.45 | 0.49 | 141 |
| LaCrO$_3$[1] | 0.323 | $Pnma$(#62.441) | 3.00 | 8 | 2.88 | 2.51 | 0.46 | 290 |
| LaCrO$_3$[2] | 0.417 | Pn'ma'(#62.448) | 2.70 | 8 | 2.88 | 3.00 | 0.40 | 310 |
| CeFeO$_3$ | 0.758 | Pnma(#62.441) | 2.60 | 8 | 4.40 | 4.05 | 0.32 | 720 |
| LaCrO$_3$[3] | 0.416 | R-3c(#167.103) | 2.80 | 8 | 2.88 | 3.00 | 0.32 | 380 |
| La$_2$NiO4 | 0.45 | Pc'c'n(#56.369) | 0.70 | 8 | 1.80 | 1.68 | 0.30 | 80 |
| FeF$_3$[1] | 0.335 | C2'/c'(#15.89) | 5.00 | 10 | 4.70 | 4.60 | 0.25 | >300 |
| FeF$_3$[2] | 0.581 | C2'/c'(#15.89) | 5.00 | 8 | 4.70 | 4.45 | 0.23 | 394 |
| CuFePO$_5$ | 0.260 | Pnma(#62.441) | 2.00 | 8 | [4.56, 0.86] | [4.28, 0.95] | 0.23 | 195 |
| CoFePO$_5$ | 0.262 | Pnm'a'(#62.447) | 1.70 | 8 | [4.54, 2.84] | [4.22, 3.36] | 0.20 | 175 |
| NaOsO$_3$ | 0.25 | Pn'ma'(#62.448) | 1.10 | 5 | 2.12 | 1.00 | 0.18 | 411 |
| Sr$_3$LiRuO$_6$ | 0.361 | C2'/c'(#15.89) | 1.80 | 6 | 2.48 | 2.03 | 0.18 | 90 |
| NiFePO$_5$ | 0.261 | Pnma(#62.441) | 2.00 | 8 | [4.54, 1.74] | [4.09, 2.03] | 0.16 | 178 |
| Sr$_3$NaRuO$_6$ | 0.404 | C2'/c'(#15.89) | 1.70 | 6 | 2.30 | / | 0.15 | 70 |
| TbFeO$_3$ | 0.351 | Pn'ma'(#62.448) | 2.30 | 8 | 4.48 | 4.80 | 0.14 | 681 |
| Ca$_3$LiRuO$_6$ | 0.239 | C2'/c'(#15.89) | 2.00 | 8 | 2.48 | 2.80 | 0.13 | 117 |
| Fe$_2$Mo$_3$O$_8$ | 0.331 | P6$_3$'m'c(#186.205) | 0.20 | 8 | [3.60, 4.30] | [4.60, 4.60] | 0.13 | 59 |
| Fe$_2$O$_3$ | 0.65 | C2'/c'(#15.89) | 2.20 | 8 | 4.50 | 4.12 | 0.10 | 955 |
| CrVO$_4$ | 1.522 | P-1(#2.4) | 2.30 | 8 | [2.90, 0] | [2.25, 0] | 0.08 | 50 |
| Ca$_3$LiOsO$_6$ | 0.3 | C2'/c'(#15.89) | 2.10 | 6 | 2.46 | 2.20 | 0.08 | 117 |
| CuFeS$_2$ | 0.802 | I-42d(#122.333) | 0.10 | 8 | [0, 3.76] | [0, 3.85] | 0.06 | 823 |

Next, we focus on two metallic altermagnets with large spin-band splitting that have not been reported before: CrSe, and CaFe$_4$Al$_8$, and one semiconducting altermagnet with extremely simple Fermi surface when hole doped that should show strong anisotropy, SrFe$_4$O$_{11}$. The crystal structure and the first Brillouin zone of the metallic altermagnet CrSe are shown in **Fig. 4a-b**. It crystalizes in P31m' (#157.55) magnetic space group with the parent space group P6$_3$/mmc (#194). We considered the collinear arrangement of moments, which is a stable configuration in eDMFT in the absence of spin-orbit coupling, and shows a similar splitting of bands as the canted configuration found in the experiment. CrSe is an example of an altermagnet from a ferrimagnetic space group, in which the vanishing of the magnetic moment is not protected by symmetry. Nevertheless, its spectra in its collinear configuration is equivalent to other altermagnets. The spectral function (**Fig. 4c**) shows a large band splitting between the *H-K-L* path. The fermi surface results show that its two different yz planes have strongly spin-split bands, while the xz and xy planes are degenerate, resulting in strong altermagnetism (**Fig. 4d**). In addition, the spectral function shows much more incoherent quasiparticle metallic states, suggesting a stronger role of correlations and possible bad metallic behavior, and likely rich phase diagram at low temperature. The crystal structure and the first Brillouin zone of the metallic altermagnet $CaFe_4Al_8$ are shown in **Fig. 4e-f**. It corresponds to the I4'/mmm' magnetic space group (139.535) with magnetic moments on Fe sites pointing approximately in the z direction. In this space group, 16 out of 32 symmetry operations flip the spins and preserve magnetic configuration, hence protecting the vanishing of the magnetic moment. The spectral function (**Fig. 4g**) shows metallic behaviour with large band splitting along *X-Γ* and *P-X-Z* paths. The fermi surface across the cuts marked in the first Brillouin zone, including the xy plane and its vertical planes, are shown in (**Fig. 4h**). The xy plane shows d-wave symmetry with notes along the marked x and y axis. The vertical planes (blue and orange planes) with 90° rotation show the opposite splitting, indicating the anisotropy of spin-band splitting of an altermagnet.

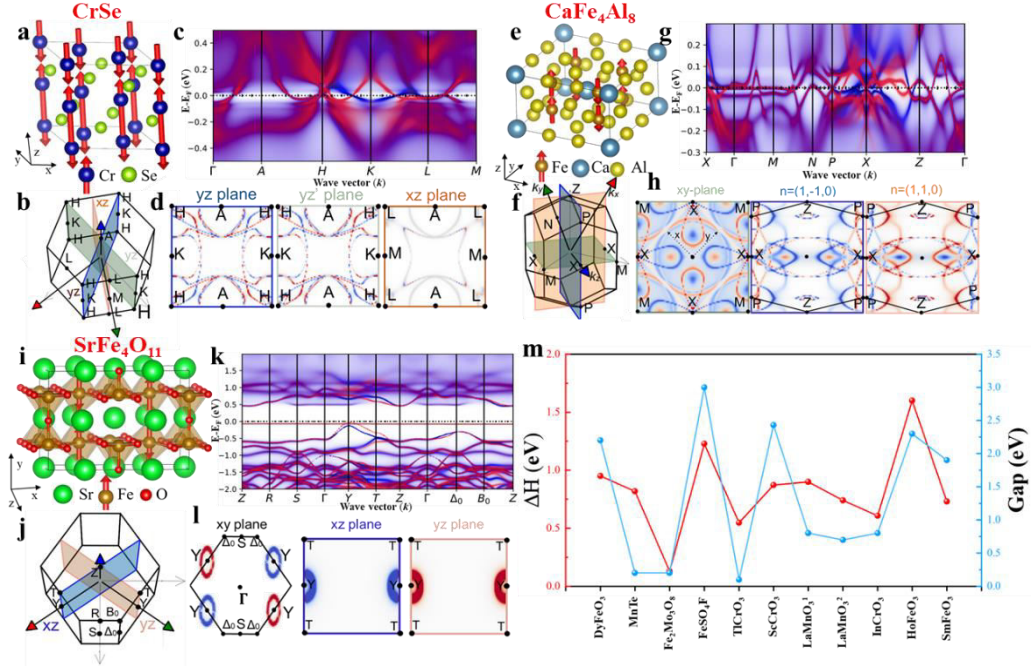

**Figure 4 The promising altermagnets identified by high-throughput screening. a,** the crystal structure of CrSe. **b,** the first Brillouin zone path of CrSe. **c,** the spectral function along the high-symmetry path of CrSe by eDMFT calculations. **d,** the fermi surface across the cuts marked in the first Brillouin zone. **e,** the crystal structure of CaFe$_4$Al$_8$. **f,** the first Brillouin zone path of CaFe$_4$Al$_8$. **g,** the spectral function along the high-symmetry path of CaFe$_4$Al$_8$ by eDMFT calculations. **h,** the fermi surface across the cuts marked in the first Brillouin zone. **i,** the crystal structure of SrFe$_4$O$_{11}$. **j,** the first Brillouin zone path of SrFe$_4$O$_{11}$. **k,** the spectral function along the high-symmetry path of SrFe$_4$O$_{11}$ by eDMFT calculations. **l,** the spectral function along the high-symmetry path of SrFe$_4$O$_{11}$ by eDMFT calculations. **m,** the band gap and ΔH of promising semiconducting altermagnets.

The structure of Sr$_4$Fe$_4$O$_{11}$ is displayed in **Fig. 4i**. It corresponds to Cmm'm' magnetic space group (#65.486) in which the centrally coordinated Fe atoms in octahedra carry larger magnetic moment (4μ$_B$) than those in square pyramids (3.6 μ$_B$). The Brillouin zone is parallelepiped (**Fig. 4j**) and the spectral function (**Fig. 4k**) along the high symmetry path in BZ shows significant spin splitting near *Y* and *T* points. Sr$_4$Fe$_4$O$_{11}$ has a 0.4eV semiconducting gap, but when hole doped, a very simple fermi surface with pockets around Y points appear, which have opposite spin orientation along the two crystallographic axes (x and y) (**Fig. 4l**), which should lead to strongly anisotropic spin transport along the two directions, demonstrating potential in spintronic applications.

In addition to $Sr_4Fe_4O_{11}$, eleven other insulating materials are identified that possess significant band splitting or narrow bandgap, including $DyFeO_3$, MnTe, $Fe_2Mo_3O_8$, $FeSO_4F$, $TlCrO_3$, $ScCrO_3$, $LaMnO_3$[1], $LaMnO_3$[2], $InCrO_3$, $HoFeO_3$, and $SmFeO_3$. Their band gaps and ΔH are illustrated in **Fig.4m**. The spectra of these altermagnets, along with the experimentally reported MnTe and four ferrimagnet candidates, can be found in **Supplementary Figure 2-8**. The details of these materials are listed in **Supplementary Note 3**. Among these semiconductors, one recent study has shown that MnTe exhibits significant altermagnetic Kramers spin splitting,[12] with the eDMFT calculations and experimental results consistent with each other, as demonstrated in **Supplementary Figure 2**. It further corroborates the effectiveness of the high-throughput screening process employed in this work. Additionally, $Fe_2Mo_3O_8$, $TlCrO_3$, and $LaMnO_3$ display narrow bandgap semiconducting characteristics and large spin-splitting, making them promising candidates for semiconductor spintronic devices.

## DISCUSSION

Benefitting from its unique magnetic property and electronic structure, altermagnet-a novel type of magnet exhibiting characteristics of both antiferromagnet and ferromagnet-shows immense potential for applications. To efficiently explore possible compositional space and discover new metallic altermagnets with significant spin-band splitting, advanced design strategies and efficient screening workflow are essential. Although recent studies, combining ARPES measurements with DFT calculations, have identified several new altermagnets based on traditional trial-and-error strategy, this approach is highly inefficient. Advanced high-throughput calculations screening offers great promise for accelerating altermagnet discovery but faces two significant challenges: reliable computational method and efficient screening processes.

While DFT + U calculations have been successfully applied to identify and study altermagnets,[35] many materials contain transition metal ions with strong local electronic correlations. Although DFT + U improves upon standard DFT by adding strong Coulomb interaction on magnetic ions and solving the atomic problem by Hartree-Fock approximation, it alleviates some of the challenges associated with strongly correlated systems, and it remains limited, particularly in predicting metallicity and band splitting in these materials. In contrast, the DFT with embedded DMFT solves the atomic problem on magnetic ions exactly, and offers significant advantages in targeting thermal state, allowing only physically relevant spin structures to be stable, avoiding predictions of completely wrong spin configurations, which are commonly found by DFT+U. In addition, the predictions of band gaps and spin-splitting are expected to be more accurate. Therefore, this work employs the eDMFT framework, which combines semi-local DFT approximation for non-magnetic ions and DMFT's dynamic treatment of site-local correlations on magnetic ions, ensuring reliable predictions of altermagnetic properties.

On the other hand, eDMFT calculations are computationally intensive. Applying eDMFT-based HTC to all candidate materials would demand immense computational resources. In this context, this work developed an automated HTC workflow that includes a pre-screening algorithm to eliminate materials unlikely to meet altermagnet requirements or future application standards, ensuring that resources are concentrated only on feasible candidates. This workflow also automates the symmetry analysis and the generation of input files needed for eDMFT calculations.

Finally, this HTC screening workflow identified four metallic altermagnets with significant band splitting from over 2,000 experimentally verified magnetic materials: CrSb, CrSe, $CaFe_4Al_8$, and $RuO_2$, along with more than ten semiconductor altermagnets with narrow bandgap and large band splitting. Many promising altermagnets identified by this method-especially metallic materials like CrSe and $CaFe_4Al_8$-were overlooked by conventional DFT+U approaches[35]. This study demonstrates that the proposed HTC screening strategy significantly accelerates the discovery of new altermagnet, minimizing both human effort and computational costs. The workflow offers a powerful framework for finding promising altermagnets and showcases the potential of applying it as a general-purpose tool for efficiently exploring broader chemical and structural spaces, driving progress in this emerging field.

## METHODS

**The eDMFT calculations**

The DFT+DMFT[18-20] calculations are performed by the eDMFT algorithm,[24,25]

which extremizes the well-defined many body Luttinger-Ward functional defined in real space. The standard DFT exchange-correlation functional is augmented by the DMFT functional. The latter requires so-called DMFT local Green's function obtained by projector, which is defined through the quasi-atomic orbitals, namely, the correlated 3d or 4d orbital on magnetic ion, with the radial dependence corresponding to the solution of the Schrödinger equation at the Fermi energy. The implementation of the algorithm starts by calculating the eigen-energies and eigen-wavefunctions of the crystal by solving the DFT equations. Next, a subset of correlated orbitals are projected out as 'quantum impurities' by the above described real-space projectors without downfolding, while the uncorrelated orbitals are treated within DFT and act as a mean-field bath on the quantum impurities, resulting in a hybridization between the two. The DMFT hybridization functions are determined self-consistently by solving the DMFT equations. The quantum impurities are solved using the hybridization expansion continuous-time quantum Monte Carlo (CTQMC) method.[36] The modified charge density obtained from the combined DFT and DMFT equations serves as the input for the next DFT-part of the iteration. The algorithm iterates until convergence is achieved for the charge density, impurity self-energies, lattice Green's function, and the free energy. Once convergence is reached, the maximum entropy method is employed for the analytical continuation from the Matsubara to the real frequency axis for both the Green's function and the self-energy.[32] The linear augmented plane wave method is used as a basis, as implemented in the WIEN2K package,[37] with the local density approximation (LDA) employed for the exchange-correlation functional.[38] We also subtract the exact double-counting between the LDA and DMFT part of the functional.[34] At each DMFT step we fully converge the charge density of the system to accuracy $10^{-5}$ to obtain very accurate impurity levels for quantum impurity models. We than solve the quantum impurity problems with at least 1Billion Monte Carlo steps and we accumulate the two-particle vertex function of two times, and the single-particle Green's function, and compute the self-energy as the ratio of the two in imaginary frequency. The imaginary time quantities are sampled using 35 SVD basis functions. The high frequency tail of the self-energy are analytically determined. The Coulomb interaction U values are estimated by constrained DMFT calculations to be of the order of 8-10 eV for 3d transition metal ions in insulators, and 6-8 eV for the same ions in metallic state. For 4d transition metals the values are typically 1-2 eV smaller, and in 5d yet another 1-2 eV smaller. The values we choose for high-throughput screening are 8 eV for 3d insulating compounds and 6 eV for 3d metallic compounds and are listed in **Table 1**. Hund's coupling is set to J=0.8 eV. The results in eDMFT are not very sensitive to the value of U, hence our calculations do not require fine tuning of these values.

**High-throughput screening**

The Magnetic Crystallographic Information files are processed by *pymatgen*.[30] Due to the presence of canted magnetic moments in many materials we simulate approximate collinear magnetic configuration to determine if altermagnetism is present. If the direction between the two magnetic moments is such that cos(theta)>0.75 or

cos(theta)<-0.75, we align the moments to be parallel or antiparallel. The equivalence of atoms is than determined by testing the shells of neighboring atoms in a radius of at least 4.7 Å. High-symmetry *k*-paths are generated by *pymatgen*, and the structures from symmetry analysis are validated by the *findsym* algorithm.[39] The crystal structure is automatically converted from standard cif configuration to wien2k structure file, including conversion to the subset of settings which are implemented in wien2k, and conversion of other settings in monoclinic and hexagonal symmetry to those that are recognized in wien2k. The optimal local coordinate axes for each correlated atom are selected by identifying the polyhedral formed by neighboring atoms to minimize the fermionic sign-problem in quantum Monte Carlo. This implementation supports various polyhedral geometries, including the cube, octahedron, tetrahedron, square pyramid, cuboctahedron, truncated tetrahedron, peak-of-tetrahedron, peak-of-square-pyramid, peak-of-hexagonal-pyramid, trigonal prism, and planar quadrilateral. The t-SNE data analysis is performed by *scikit-learn* package.[40]

**Declaration of Competing Interest**

The authors declare that they have no known competing financial interests or personal relationships that could have appeared to influence the work reported in this paper.

**Data availability**

Data used in this work are available via internet (https://www.cryst.ehu.es/magndata/).

**Code availability**

The code of the eDMFT can be found at http://hauleweb.rutgers.edu/tutorials/. The code of the data analysis and high-throughput screening is available at https://github.com/XuhaoWan/HT_AM.

**Note 1.** The input of all magnetic materials in MAGNDATA database in t-SNE analysis.

The specific input, from left to right, consists of the following: ID (1), one-hot encoding of the elements present (103), the specific proportions of the elements (103), one-hot encoding of the crystal system (8), the total number of propagation vectors (1), the unit cell volume (1), and the lattice constants *a*, *b*, *c* (3), α, β, and γ (3).

The numbers in parentheses represent the total number of columns for each property. For elements, columns corresponding to all 103 elements from H to Lr were initially generated. Subsequently, during the t-SNE analysis, columns corresponding to elements that had a value of 0 across all over 2,000 magnetic materials were removed to eliminate redundant features and improve the clarity of the analysis. The crystal system includes seven categories: triclinic, monoclinic, orthorhombic, tetragonal, trigonal, hexagonal, and cubic, along with unknown, representing cases where no match was found through analysis using *Pymatgen*. Before performing the t-SNE analysis, the generated matrix will be normalized column-wise, following the normalization formula:

$$x' = \frac{x - x^{min}}{x^{max} - x^{min}} \quad (S1)$$

Where $x'$ is the original data, $x'$ is the value after normalization, $x^{max}$ and $x^{min}$ is the maximum and minimum value in each column, respectively.

**Note 2.** The detailed method to calculate the measure of spin splitting.

The measure of spin splitting ($\Delta H$) is calculated from the eDMFT spectral function, representing the splitting between the spin-up and spin-down states, using the following formula:

$$\Delta H = \frac{1}{N_k} \sum_{N_k} \int_{E_{min}}^{E_{max}} |Tr(A_\uparrow(\boldsymbol{k},\omega) - A_\downarrow(\boldsymbol{k},\omega))\omega \, d\omega| \qquad (S2)$$

Here, we take the absolute value of the average over momentum points because the average without the absolute value vanishes in altermagnets, as protected by symmetry operations. In this momentum sum we also filter out contributions from k-points where splitting vanishes due to symmetry. We compute the first frequency moment of the spectral function, which has the units of energy. This quantity has a simple form in the non-interacting limit, where $As_{ij}(\boldsymbol{k},\omega) = \delta_{ij}\delta(\omega - \epsilon_{i,\boldsymbol{k},s})$, yielding:

$$\Delta H_{non-interacting} = \frac{1}{N_k} \sum_{N_k} \left| \sum_{E_{min} < \epsilon < E_{max}} \epsilon_{\uparrow i,\boldsymbol{k}} - \epsilon_{\downarrow i,\boldsymbol{k}} \right| \qquad (S3)$$

This expression clearly measures the average value of the spin splitting between the spin-up and spin-down bands. In our calculation, we take the energy window between $E_{min} = -3 \, eV$ and $E_{max} = 3 \, eV$. The spectral function in the equation above is computed as:

$$A_s(\boldsymbol{k},\omega) = \frac{1}{2\pi i}\left(G_s^\dagger(\boldsymbol{k},\omega) - G_s(\boldsymbol{k},\omega)\right) * (-\frac{1}{\pi}) \qquad (S4)$$

where the Green's function is a matrix expressed in the complete Kohn-Sham basis:

$$G_s(\boldsymbol{k},\omega) = (\omega + \mu - \varepsilon(\boldsymbol{k}) - \Sigma_s(\boldsymbol{k},\omega))^{-1} \qquad (S5)$$

and $\varepsilon(\boldsymbol{k})$ are the Kohn-Sham energy bands, given by $\varepsilon_{ij}(\boldsymbol{k}) = \delta_{ij}\epsilon_{i,\boldsymbol{k}}$, computed on the eDMFT charge density. The term $\Sigma_s(\boldsymbol{k},\omega)$ represents the embedded DMFT self-energy:

$$\Sigma_{s_{ij}}(\boldsymbol{k},\omega) = \sum_{\alpha,\beta}\langle\psi_{i,\boldsymbol{k}}|\phi_\alpha\rangle\Sigma_{\alpha,\beta}(\omega)\langle\phi_\beta|\psi_{i,\boldsymbol{k}}\rangle \qquad (S6)$$

where $\psi_{i,\boldsymbol{k}}$ are Kohn-Sham eigenvectors, and $\phi_\alpha$ are quasi-atomic orbitals, in which DMFT selfenergy ($\Sigma_{\alpha,\beta}(\omega)$) is computed. Note that this self-energy is the quantum impurity self-energy corrected for double-counting self-energy.

**Note 3.** The detailed information of materials shown in Supplementary figures.

The Brillouin zone paths along with the eDMFT spectral function of the ten semiconducting altermagnets including $DyFeO_3$, $MnTe$, $Fe_2Mo_3O_8$, $FeSO_4F$, $TlCrO_3$, $ScCrO_3$, $LaMnO_3^1$, $LaMnO_3^2$, $InCrO_3$, $HoFeO_3$, are presented in **Supplementary Figure 3** ($Fe_2Mo_3O_8$, $FeSO_4F$), **Supplementary Figure 4** ($TlCrO_3$, $ScCrO_3$, $SmFeO_3$), **Supplementary Figure 5** ($LaMnO_3^1$, $LaMnO_3^2$, $HoFeO_3$), and **Supplementary Figure 6** ($InCrO_3$, $DyFeO_3$). The MnTe exhibits significant altermagnetic Kramers spin splitting, with the eDMFT calculations and experimental results consistent with each other, as demonstrated in **Supplementary Figure 2**. In short, $Fe_2Mo_3O_8$, $TlCrO_3$, and $LaMnO_3$ display narrow bandgap semiconducting characteristics and large spin-splitting, making them promising candidates for semiconductor spintronic devices. Besides, the eDMFT spectral function of four candidates not strictly altermagnets but weak ferrimagnetic with finite splitting at $\Gamma$ point determined by HTC including $Cr_2S_3$, $CaFe_2O_4$, $BiCrO_3^1$, and $BiCrO_3^2$ are shown in **Supplementary Figure 7-8**.

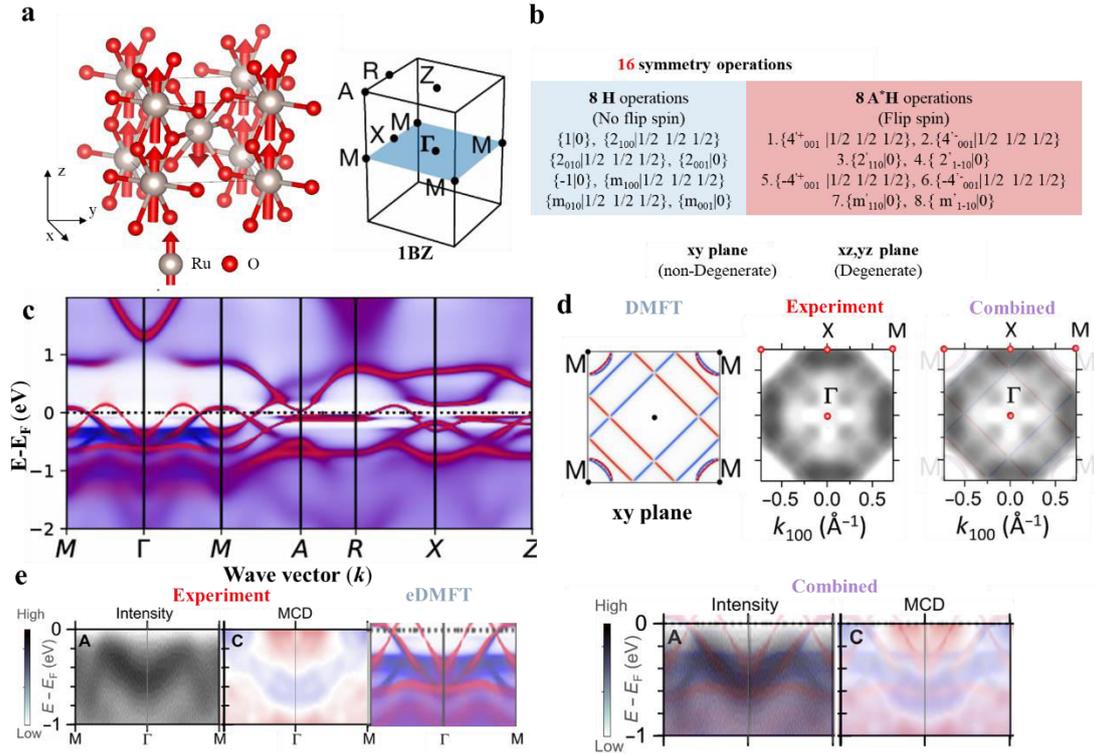

**Figure 1. a,** the crystal structure and first Brillouin zone of $RuO_2$. **b,** The symmetry operation analysis of $RuO_2$ as an altermagnet. **c,** the spectral function of $RuO_2$ along the high-symmetry path by eDMFT calculations. **d,** the comparison of the xy Fermi surface of $RuO_2$ with experimental photoelectron intensity at the fermi surface for the xy plane. **e,** the comparison of the spectral function of $RuO_2$ with the experimental measured intensity map and magnetic circular dichroism (MCD) results.[1]

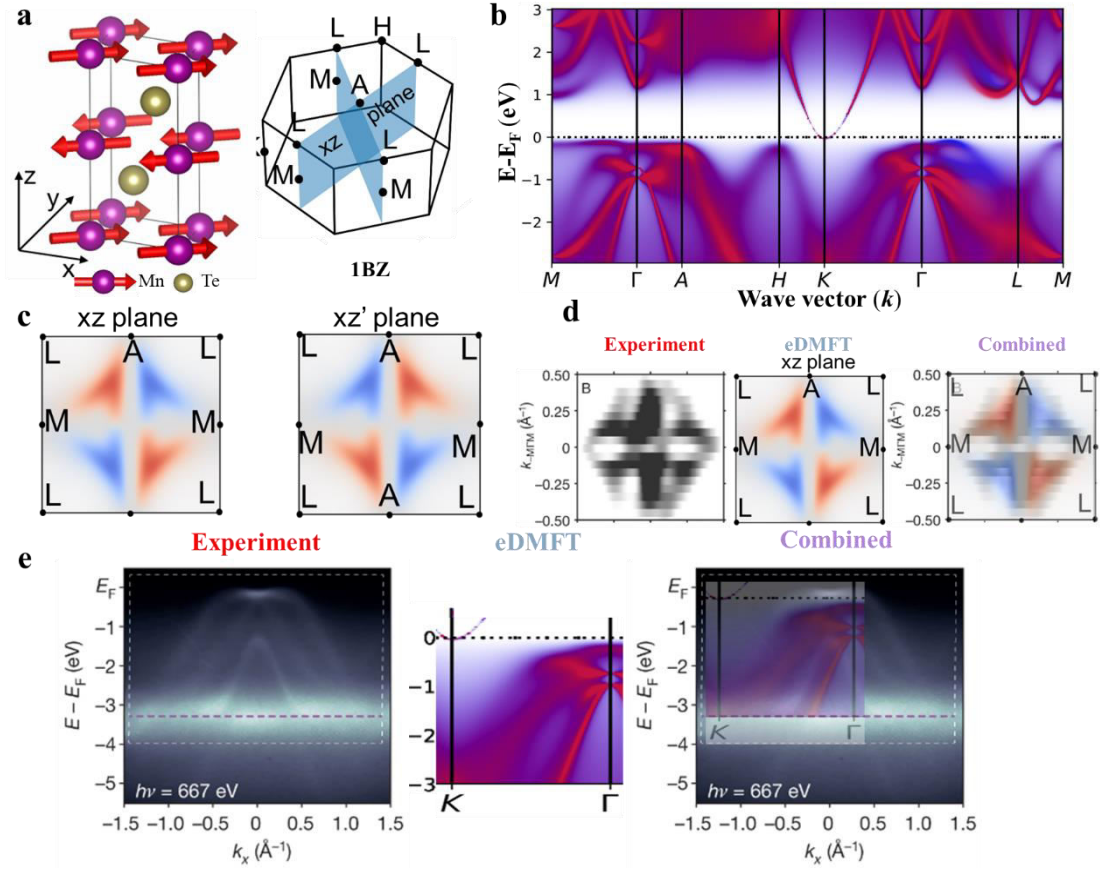

**Figure 2. a,** the crystal structure and first Brillouin zone of MnTe. **b,** the spectral function of MnTe along the high-symmetry path by eDMFT calculations. **c,** the xz fermi surface of MnTe which should be non-degenerate by symmetry analysis. **d,** the comparison of the xz fermi surface of MnTe with experimental measured constant-energy maps for binding energies of the xz fermi surface. **e,** the comparison of the spectral function of MnTe with the experimental soft X-ray (667 eV) ARPES band map at $\Gamma$-$K$ path.[2]

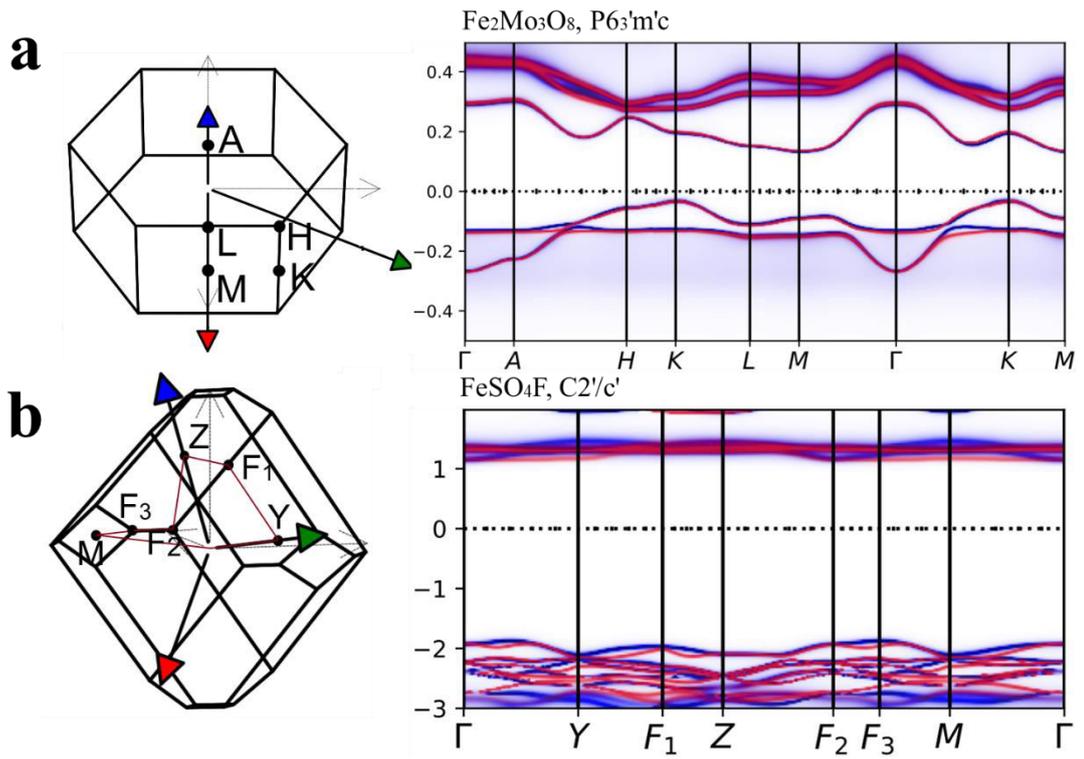

**Figure 3.** the first Brillouin zone and the spectral function along the high-symmetry path by eDMFT calculations of **a,** Fe$_2$Mo$_3$O$_8$, and **b,** FeSO$_4$F.

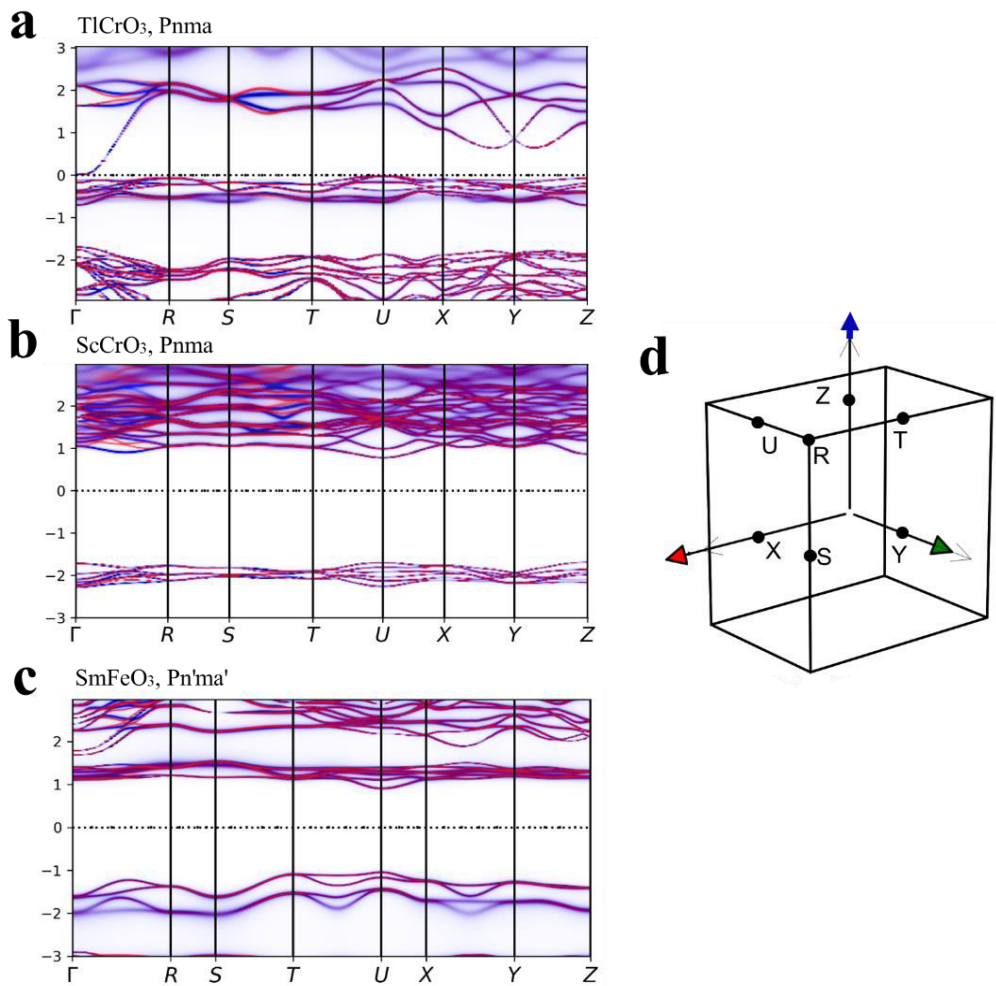

**Figure 4.** The spectral function along the high-symmetry path by eDMFT calculations of **a,** TlCrO$_3$, **b,** ScCrO$_3$, and **c,** SmFeO$_3$. **d,** the first Brillouin zone of the three materials

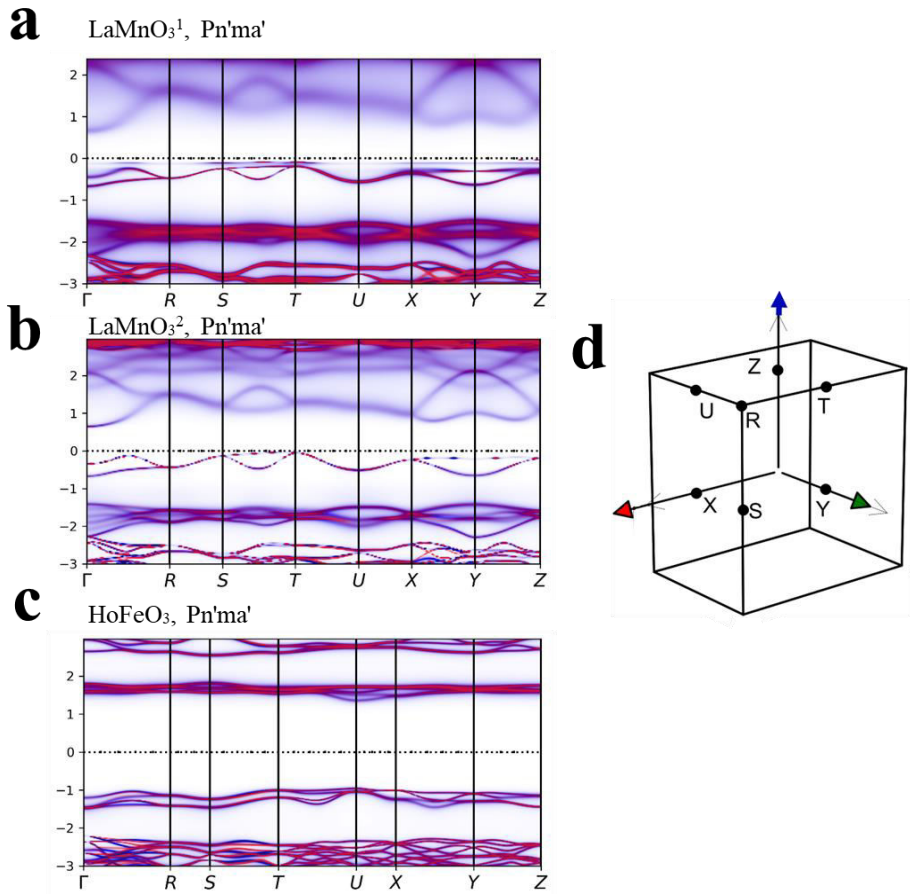

**Figure 5.** The spectral function along the high-symmetry path by eDMFT calculations of **a,** LaMnO$_3^1$, **b,** LaMnO$_3^2$, and **c,** HoFeO$_3$. **d,** the first Brillouin zone of the three materials

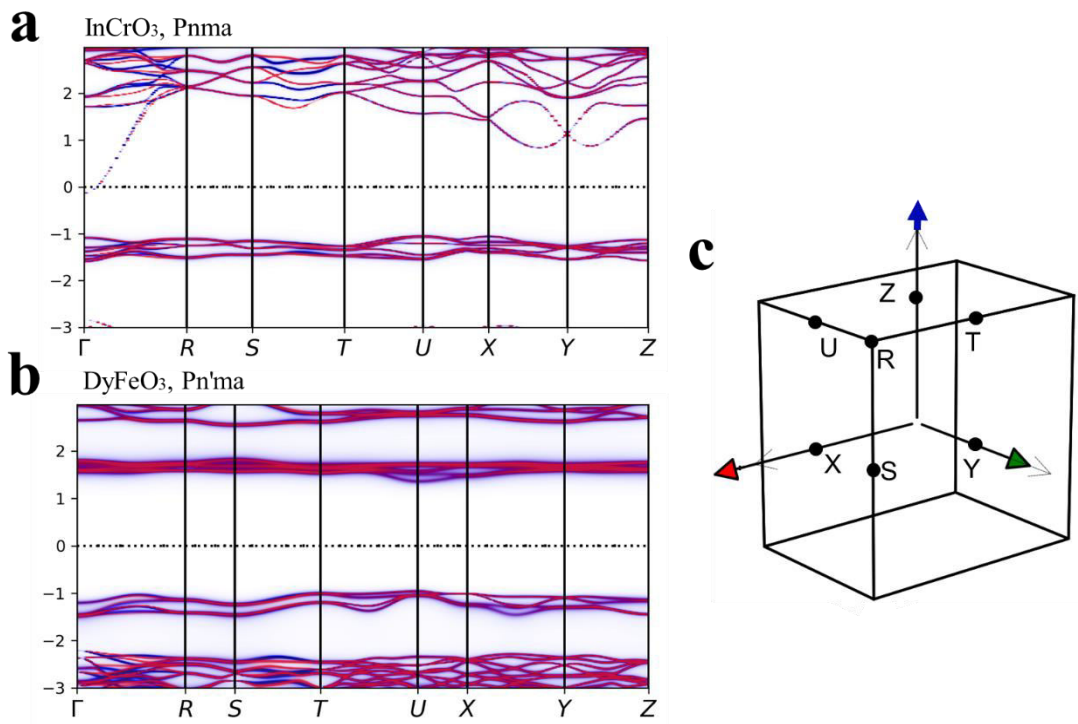

**Figure 6.** The spectral function along the high-symmetry path by eDMFT calculations of **a,** InCrO$_3$, **b,** DyFeO$_3$, and **c,** the first Brillouin zone of the two materials

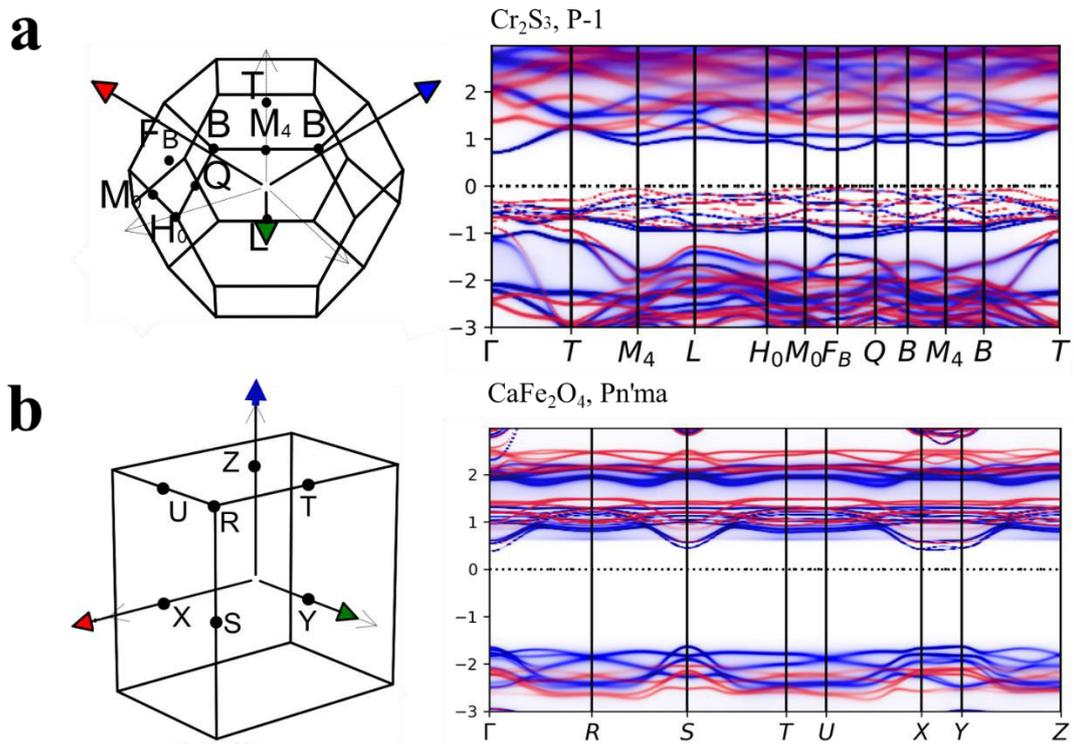

**Figure 7.** the first Brillouin zone and the spectral function along the high-symmetry path by eDMFT calculations of **a,** Ferrimagnet $Cr_2S_3$, and **b,** Ferrimagnet $CaFe_2O_4$.

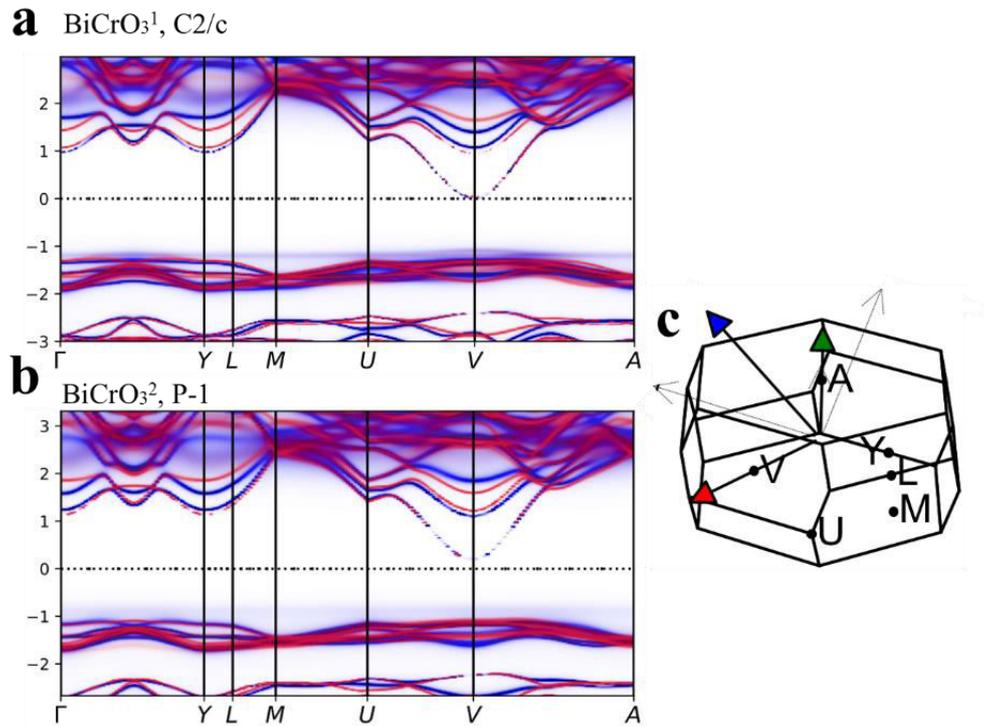

**Figure 8.** The spectral function along the high-symmetry path by eDMFT calculations of **a,** Ferrimagnet $BiCrO_3^1$ (ID: 0.138),and **b,** Ferrimagnet $BiCrO_3^2$ (ID: 0.139) **c,** the first Brillouin zone of the two $BiCrO_3$.